# EdgeLens: Deep Learning based Object Detection in Integrated IoT, Fog and Cloud Computing Environments

Shreshth Tuli[1,2], Nipam Basumatary[1,3] and Rajkumar Buyya[1]
[1]CLOUDS Laboratory, School of Computing and Information Systems, The University of Melbourne, Australia
[2]Department of Computer Science, Indian Institute of Technology (IIT) Delhi, India
[3]Department of Computer Science, Indian Institute of Technology (IIT) Madras, India

*Abstract*— Data-intensive applications are growing at an increasing rate and there is a growing need to solve scalability and high-performance issues in them. By the advent of Cloud computing paradigm, it became possible to harness remote resources to build and deploy these applications. In recent years, new set of applications and services based on Internet of Things (IoT) paradigm, require to process large amount of data in very less time. Among them surveillance and object detection have gained prime importance, but cloud is unable to bring down the network latencies to meet the response time requirements. This problem is solved by Fog computing which harnesses resources in the edge of the network along with remote cloud resources as required. However, there is still a lack of frameworks that are successfully able to integrate sophisticated software and applications, especially deep learning, with fog and cloud computing environments. In this work, we propose a framework to deploy deep learning-based applications in fog-cloud environments to harness edge and cloud resources to provide better service quality for such applications. Our proposed framework, called EdgeLens, adapts to the application or user requirements to provide high accuracy or low latency modes of services. We also tested the performance of the software in terms of accuracy, response time, jitter, network bandwidth and power consumption and show how EdgeLens adapts to different service requirements.

## I. INTRODUCTION

The cloud computing paradigm has provided users the "pay-as-you-go" computation service which is an efficient alternative to owning and managing private data centers both for individuals and industries [1]. This Internet-based remote utility computing model becomes a problem for various latency-sensitive applications like healthcare and surveillance which require real time results [2]. The emerging wave of IoT deployments are quickly shifting to application or services that are both resource hungry as well as require low response times. To solve this, the Fog computing paradigm has emerged which harnesses edge resources along with remote cloud resources as required [3]. Fog computing has gained significant importance due to its robustness and the ability to provide diverse response characteristics based on target applications. This new paradigm has facilitated and enhanced the mobility, privacy, security and latency of real-time applications [4].

Object detection and surveillance have become pervasive in modern digital society [5] with ubiquitous deployment of IoT enabled devices such as cameras everywhere. Autonomous video surveillance is a process of analyzing video sequences using object detection, segmentation and classification for various applications. Thus, object detection being a critical part, is an active area of research in computer vision. In the last few years there have been significant advances in the field of computer vision and deep learning which have made such object detection software much faster, accurate and precise. When such systems are deployed in real-life, they are still unable to perform well due to increasing data rates and poor system frameworks. Even with efficient models that are able to perform high resolution and high frame-per-second video analysis in very less time, the network overhead and transfer times of huge volumes of data renders these deployments useless. Fog computing seems to be a good solution to bring down the data transfer time significantly and hence improve the responsiveness and quality of such applications.

There have been some works to adopt such deep learning applications to fog environments [6] but none of them provide framework for seamless integration of IoT-Fog-Cloud which can be deployed with engineering simplicity and adapt to the user and application requirements. The lack of such models or frameworks that integrate the power of high accuracy of deep learning models with the responsiveness of edge nodes motivated this work.

In this work, we have developed a deep learning-based fog-cloud deployable system, called *EdgeLens*, for real time object detection. Our framework uses Aneka service [7] and adapts to user and/or application needs to provide two modes in which the results are generated in high accuracy mode and other in low latency mode. The key **contributions** of this paper are:

- A generic system architecture for development of deep learning applications in fog and cloud computing environments

- A lightweight automatic object detection system using deep learning and Aneka which is called EdgeLens

- Demonstrated and analyzed EdgeLens in terms of various performance parameters like accuracy, response time, network bandwidth and power consumption.

The rest of the paper is organized as follows. Section II presents related work towards deep learning – fog integration frameworks. Section III proves the background details of Aneka platform which was used to develop and deploy the framework. Section IV gives the design and architecture of EdgeLens with implementation in Section V. Section VI describes the experimental setup and the performance evaluation of the framework. Conclusions and future work are proposed in Section VII.

| Work | IoT | Fog Computing | Cloud Computing | Deep Learning | Performance Parameters ||||  |
|---|---|---|---|---|---|---|---|---|---|
| | | | | | Accuracy | Response Time | Jitter | Power | Network Bandwidth |
| Chen et al. [8] | ✓ | ✓ | ✓ | | | ✓ | | | |
| Diro et al. [9] | ✓ | ✓ | | ✓ | ✓ | ✓ | | | |
| Li et al. [10] | ✓ | ✓ | | ✓ | ✓ | | | | |
| Teera et al. [11] | | ✓ | ✓ | ✓ | | | | | ✓ |
| Constant et al. [12] | ✓ | ✓ | ✓ | ✓ | | ✓ | | ✓ | |
| **EdgeLens** | ✓ | ✓ | ✓ | ✓ | ✓ | ✓ | ✓ | ✓ | ✓ |

Table 1: A comparison of existing systems with EdgeLens

## II. RELATED WORK

Chen et al. [8] modelled a smart surveillance architecture for detecting fast moving vehicles. The video obtained is processed on fog computing nodes such as smartphones, smart tablets, computers in police cars and other onsite devices with computation capabilities. But their model is only able to detect single vehicle at a time and not multiple target objects. Diro et al. [9] implemented a distributed attack detection scheme using deep learning. They used fog nodes for training models and implemented the attack detection at the edge of the fog network. The master node performs parameter sharing and optimization. Their experiments showed that distributed attack detection using deep learning model is better than centralized attack detection system. But they did not compare their deep learning model with other traditional machine learning algorithms such as SVM, decision trees and other neural networks.

Li et al. [10] showed deep learning for smart industry with fog computing framework. Their model handles the large data obtained from sensors adopted in industrial productions to detect defects of the products by offloading the computation burden from central server to the fog nodes. They have used convolutional neural network (CNN) for the predictive analysis and also simultaneously indicated the defect type and its degree. Teerapittayanon et al. [11] proposed a distributed deep learning network (DDNN) model over a distributed computational hierarchy consisting of the cloud, the edge (fog) and the end devices. They showed that by processing more sensor data on end devices rather than offloading on the cloud, they achieved high accuracy and also reduced communication cost. But they performed all their experiments on binary Neural Network layers (NN) and didn't consider the possibility of mixed precision or floating-point NN layers. Constant et al. [12] developed a prototype of smart gateway that performed the process of data conditioning, intelligent filtering, smart analytics and selective transfer to the cloud for long term storage and temporal variability monitoring. Thus, with smart gateways they introduced end-to-end interaction between the sensor devices and the cloud.

A summary of comparison with related work is shown in Table 1. Our work provides a simplistic and lightweight distributed implementation of deep learning applications over integrated fog and cloud computing environments using IoT.

## III. BACKGROUND TECHNOLOGY - ANEKA

To develop and deploy the proposed system, we leveraged the computing capabilities of edge and cloud resources using Aneka platform [7]. Aneka is a platform for developing and deploying applications on cloud infrastructure. It provides a runtime environment and APIs that allow developments of .NET applications that harness computation capabilities of public or private clouds [7]. The public cloud can comprise of Virtual Machines (VMs) provided by cloud service providers like Azure or Amazon Web Services. The private cloud can comprise of enterprise cloud VMs, fog or edge devices in the Local Area Network (LAN). The core components of the Aneka framework are designed and implemented in a service-oriented fashion. Aneka provides dynamic provisioning which is the ability to dynamically acquire resources and integrate them into existing frameworks and software solutions. Dynamic provisioning in provided in Aneka using two main services: Resource provisioning and Scheduler service. Aneka provides multiple programming models, such as Thread, Task and Map-Reduce. The Task model, supported in Aneka, considers each job request as a task and distributed all tasks across various virtual resources available across private or public cloud. Thus, Aneka provides a seamless execution environment for efficiently integrating Edge and Cloud resources and gives a unified interface for data-intensive applications. EdgeLens uses the Aneka Task model to distribute object detection tasks across different fog and cloud resources.

## IV. SYSTEM ARCHITECURE

An IoT based fog-cloud integration system architecture for object detection which can manage input images effectively to provide results in near real-time is shown in Figure 1. It integrates different hardware and software components and allows structured communication.

### A. Components
The system has the following hardware components:

*1) Input Sensors:* These include cameras and video-cameras that may or may not be attached with the gateway device.

*2) Gateway:* Different types of gateway devices exist which include mobile phones, laptops and tablets. These act as fog devices that collect the images from cameras and forward them to the Aneka Master node for detection.

*3) Aneka Master:* This is the Aneka master container in fog environment, that takes job requests from the gateway device

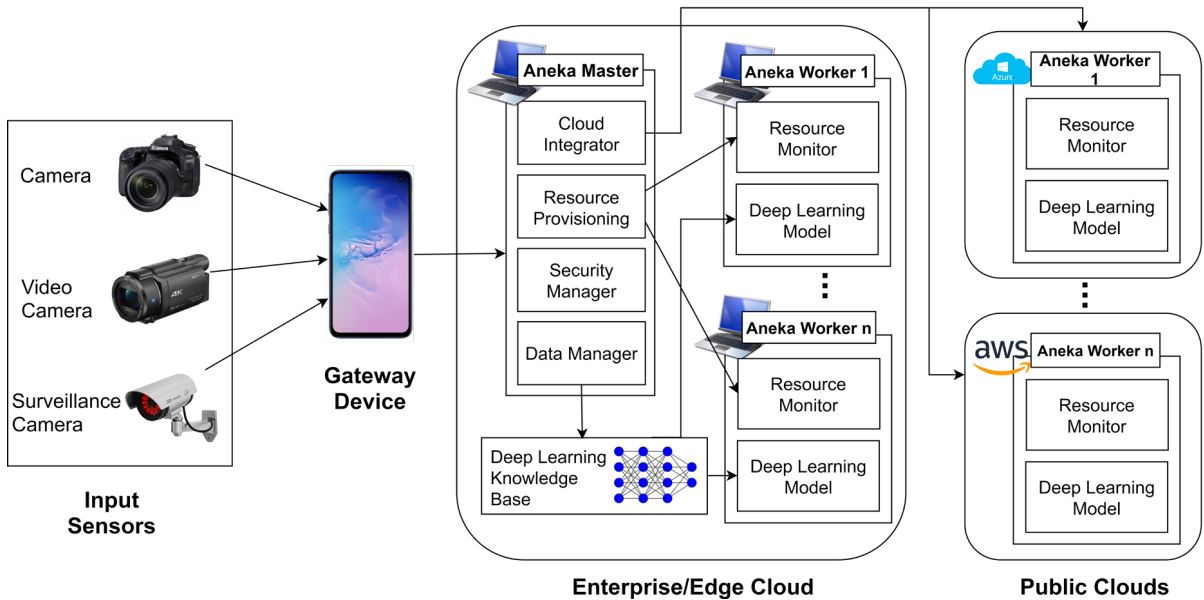
Figure 1: Architecture of proposed system

and sends it to worker containers. It contains the dynamic provisioning, load balancing and scheduling models that helps in task distribution across the Aneka containers.

*4) Aneka Worker:* The worker containers present in private cloud (fog environment) or public cloud (like Azure or Amazon Web Services) perform the computation task and comprise of the deep learning models for object detection. The fog nodes may comprise of Single Board Computers (SBCs) like Raspberry Pis. The task sent to the Aneka Master is forwarded to the Aneka Worker containers.

### B. Software Services

The software components of the proposed system are:

*1) Fabric Services:* These implement the fundamental opertions of the infrastructure of public and private clouds. These services include resource provisioning, failover mechanism for improved reliability, performance monitoring and hardware profiling.

*2) Foundation Services:* These help in enhancing the application execution in the cloud and include resource reservaion, billing, monitoring and storage management.

*3) Gateway Interface:* This is the Graphical User Interface (GUI) at the gateway device that helps in configuring the sensors and indicating the address of the Aneka Master device.

*4) Deep Learning Module:* This module performs the computation task and converting input image to resulting segmented and classified ouput image. More detailed description is given in Section V.C.

### C. Network Topology

The topology of the proposed system follows Master-Worker fashion where there is a single Aneka Master container and multiple Worker containers in private or public cloud. The edge devices including the fog worker nodes and the gateway device are present in the same Local Area Network (LAN). The Aneka Master is also in a Virtual Private Network (VPN) to harness the cloud resources which might be Virtual Machines (VM) in a Cloud Data Center (CDC).

## V. DESIGN AND IMPLEMENTATION

### A. Pre-processing

The proposed system allows users to configure the data processing in two modes: High-Accuracy mode or Low-Latency mode. Based on the mode selected, the model adapts the resource distribution and input pre-processing. In the former mode, the model sends raw input image, without compression, to the Aneka Master container. In low-latency mode, the input image is scaled down in resolution so that the network latency and execution time is lower, but may compromise on the detection accuracy. The rescaling is performed by *rescale.py* in Master.

### B. Task Parallelism

EdgeLens uses Aneka-Task model for distribution of workload across the Edge nodes and Cloud VMs. The task parallelism code was developed in C# using .NET framework. When the Aneka Master gets the input image after pre-processing, it creates an Aneka Task and sends it to one of the workers for processing. At the Aneka worker container, the task invokes the Python based deep learning software *yolo.py* (discussed in Section V-C). The results after computation through the deep learning application is fetched by Aneka Master and forwarded to the Gateway device.

### C. Deep Learning Application - You Only Look Once (YOLO)

YOLO (You Only Look Once) is a model for object detection [14] used in this framework. The task is to detect the location of objects in the image by creating bounding box about the objects and then classify them into different classes. Methods like Recurrent Convolutional Neural Network (R-CNN) and its variations used in multiple steps to perform this

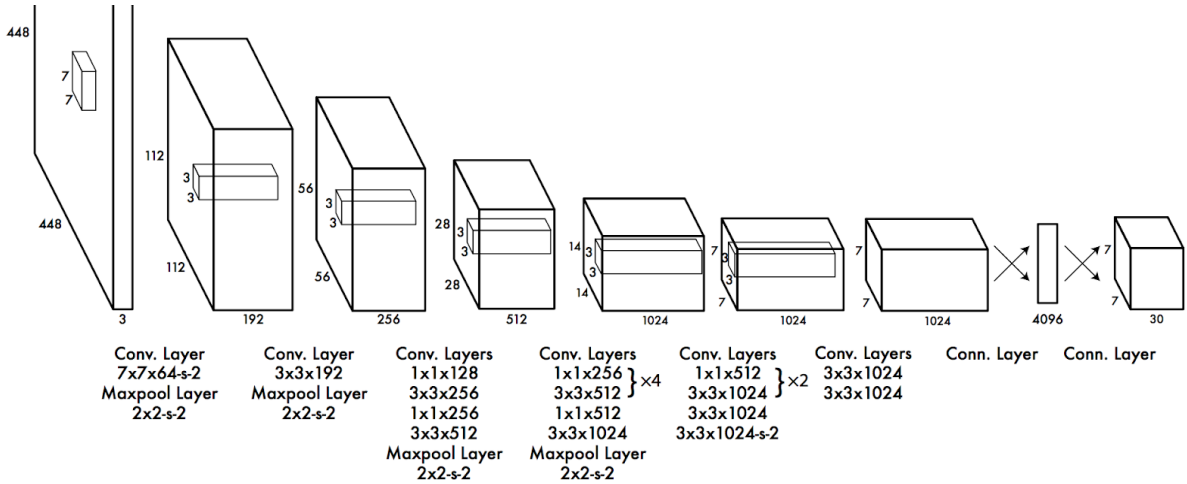
Figure 2: Neural Network architecture of YOLO

task. This turned out to be slow and each individual component had to be trained separately. But YOLO has an end-to-end architecture which does region proposal and classification simultaneously and thus provides faster results.

The image is divided into S×S grid and each cell is interested in predicting 5+k (k is number of classes) quantities which are probability (confidence) that this cell is indeed contained in a true bounding box, width and height of bounding box, center (x,y) of the bounding box and probability of the object in the bounding box belonging to the kth class (k-values). The output layer thus contains S×S×(5+k) elements. Now for each cell the bounding box, its confidence and the object in it are computed. Then the most significant bounding boxes and the corresponding object label are retained. The YOLO model, as shown in Figure 2, is implemented using CNN where the initial layers extract the features from the image and then fully connected layers predict the output probabilities and coordinates. The network has 24 convolutional layers followed by 2 fully connected layers. Alternating 1×1 convolutional layers reduce the features space from preceding layers.

The YOLO model is extremely fast and simple [14]. It does not require a complex pipeline as it detects the frame using a regression mechanism. Each grid cell predicts bounding boxes and their respective confidence scores. Each grid cell also predicts conditional class probabilities. Using both class probability map and bounding boxes with confidence score, it makes the final prediction.

### D. Android Interface

EdgeLens is built for android smartphones that act as gateway devices. The application interface is shown in Figure 3. The interface was developed using MIT's App Inventor [13]. This interface allows the user to set the network address of the node running the Aneka Master container. The communication of the input image received from the camera is achieved by Aneka web-service. This service uses HTTP POST request to send the image for computation. The Aneka Master then decides which worker to send the image and forwards it using the Aneka File Transfer Protocol (FTP).

## VI. PERFORMANCE EVALUATION

To demonstrate the feasibility and efficiency of the proposed framework, we developed and deployed it for object detection in images in Fog-Cloud integrated computing environment.

### A. Experimental Setup

The system setup for evaluation of the model is described below and shown in Figure 4:

*1) Gateway device:* Samsung Galaxy S7 with Android 9, in CLOUDS Lab at University of Melbourne, Victoria, Austalia.

*2) Aneka Master:* Dell XPS 13 with Inte® Core™ i5-7200 CPU @ 2.50GHz, 8.00 GB DDR4 RAM and 64-bit Windows 10, in CLOUDS Lab at University of Melbourne, Victoria, Austalia.

*3) Aneka Fog Worker:* Dell Lattitude 5490 with Intel® Core™ i7-8650U CPU @ 1.9GHz, 16.00 GB DDR4 RAM and 64-bit Windows 10, in CLOUDS Lab at University of Melbourne, Victoria, Austalia.

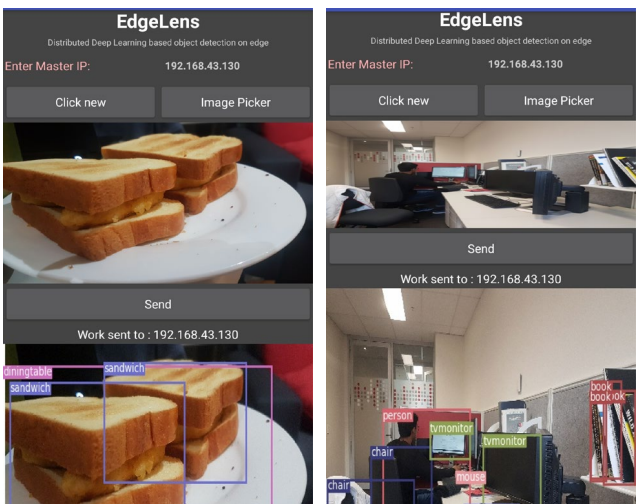
Figure 3: Android Interface at gateway device

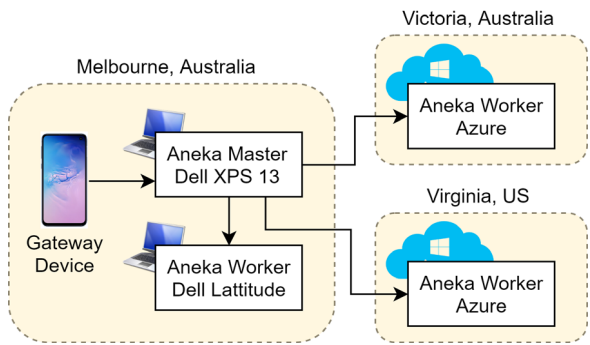

Figure 4: Experiment setup

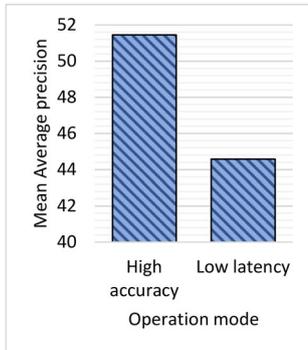

Figure 5: mAP results

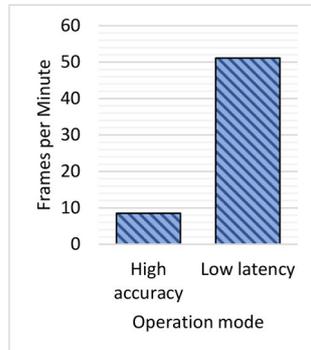

Figure 6: FPM results

*4) Aneka Cloud Worker:* Microsoft Azure B1s Machine, 1vCPU, 1GB RAM, 2GB SSD, Windows Server 2016 in two locations: Victoria, Australia and Virginia, USA.

The YOLOv3 implementation [15] was used with pre-trained weights by training on the COCO dataset [16]. The Aneka C# code invokes a Python module with this implementation and was used to run the experiments.

During the experiments, data was sent with frequency of 10 images per minute. Data parameters were recorded using Microsoft Performance Monitor [17] at the Master and the Azure VM. To measure the network bandwidth consumption Microsoft Network Monitor 3.4 [18] was used at the Aneka Master node.

### B. Detection Performance

To evaluate an object detection model, we used the mean Average Precision (mAP) metric [19]. The Average Precision (AP) is calculated as the area under the precision-recall curve and mAP is the average AP values for all detection classes. To compare the two operation modes in EdgeLens (high-accuracy and low-latency), we evaluated the model on the PASCAL-Visual Object Classes (VOC) Challenge dataset [20]. The mAP of the two modes with a single worker node is shown in Figure 5. The detection and response speed of the two modes are compared using the Frames per Minute (FPM) metric [19] which indicates the number of images they can detect in a minute is shown in Figure 6 for a single worker node. The graphs show that the high-accuracy mode has higher detection performance but is much slower than the low-latency mode.

Figure 7 shows the variation in the detection results for the two configurations of high accuracy and low latency. The input is an image of a computer lab with dim lighting and is of size 4000×2192 pixels or 978 KB. For high-accuracy result the image is considered as it is, but for low-latency result the image is scaled down to 200×110 pixels of size 4.84 KB. The high-accuracy output classifies most monitors, keyboards and chairs correctly. The low latency output has significantly poor detection result and does not detect many monitors or chairs and even incorrectly classifies some of them.

### C. Response Time

Fog different Fog scenarios, the average response times are shown in 8. We see that the response time is low when the data is sent to the fog nodes compared to when it is sent to cloud container. This is expected because the data transfer time in the LAN is much lower than the Internet to CDC. Also, for two fog worker nodes, the average response time is lower because of a greater number of resources for the tasks. For every case, we can observe that the response time is much lower when in low-latency mode compared to when in high-accuracy mode.

### D. Jitter

Jitter is variation in latency and is an important measure for many applications like surveillance. Figure 9 shows the jitter in different cases. We see that jitter is highest when tasks are sent to cloud VM and low when tasks sent to fog nodes. Also, jitter is lower when there are two edge nodes compared to the case with only one edge node. Jitter is lower for low latency mode too.

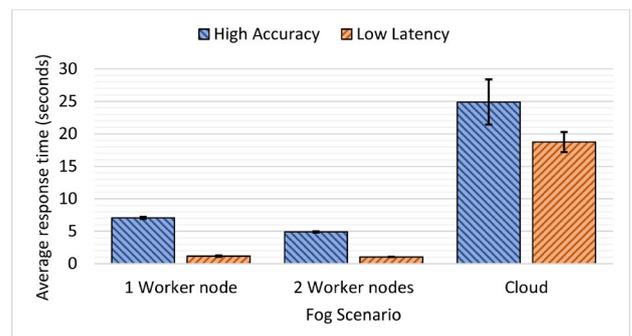

Figure 8: Response time comparison

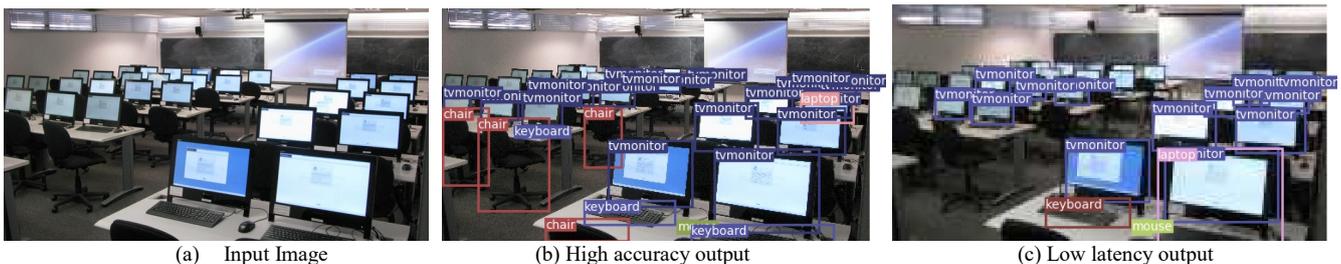

(a) Input Image  (b) High accuracy output  (c) Low latency output

Figure 7: Input image and outputs in different configurations

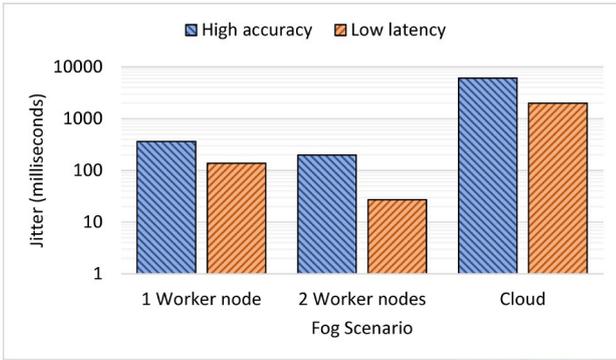

Figure 9: Jitter comparison

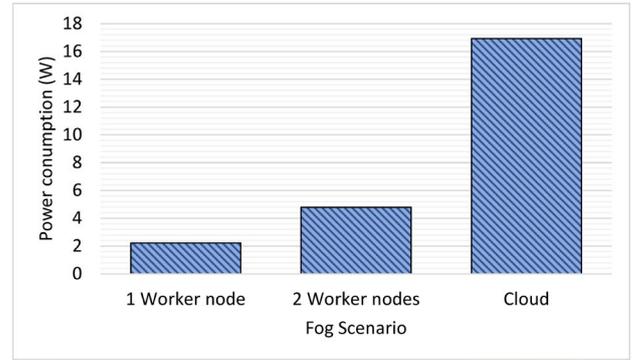

Figure 11: Power comparison

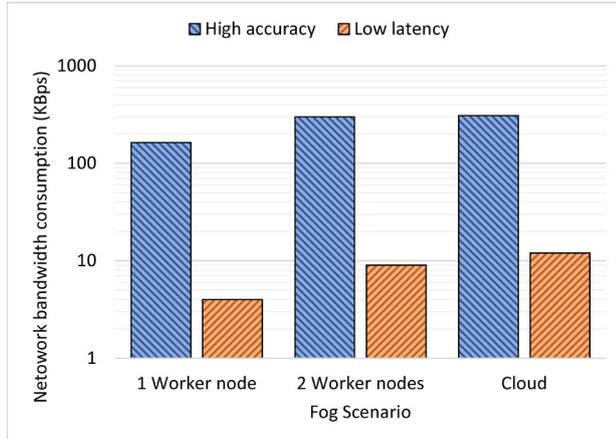

Figure 10: Network bandwidth consumption comparison

*E. Network Bandwidth Usage*

The network bandwidth consumption for different cases is shown in Figure 10. As expected, the network consumption is very high when configured to run in high-accuracy mode compared to when in low-latency mode. This is because in high-accuracy mode ten 0.9 MB images are being sent every minute but in low-latency mode ten 4.8 KB images are being sent every minute.

*F. Power Consumption*

Figure 11 shows the total power consumption of all devices in different scenarios. We see that the power consumption is highest in the case when all tasks are sent to cloud VM. Power consumption of fog nodes is very low and increases to almost double when we have two edge nodes compared to the case with only one edge node.

*G. Discussion*

Based on the experiments, we suggest that the proposed approach can be used in the following settings based on the target Quality of Service (QoS):

- For latency critical tasks that are lightweight where results are not sensitive to accuracy, the low-latency mode should be used with fog configuration.
- For tasks requiring very high accuracy but network and energy are not limited high-accuracy mode must be used. If tasks require very heavy computation then Cloud must be used.

This work demonstrates capabilities of Aneka offering a lightweight means to develop and test distributed deep learning applications using Aneka with engineering simplicity and robustness.

## VII. CONCLUSIONS AND FUTURE WORK

We proposed a novel fog-cloud based deep learning approach for object detection. Our system provides a deployable framework for deep learning applications and provides different modes (high-accuracy and low-latency) for different target applications or user requirements. We used Aneka platform service to deploy and test the effectiveness of the proposed model. We compared different characteristics like detection accuracy, response time, jitter, network and power consumption for different fog scenarios and used the results to suggest different modes of operation for different use cases.

As part of the future work, we plan to extend the system to consider a given cost model to distribute and share resources or pre-process/resize input images based on budget constraints of the user. Also, currently the training of the model is done separately on a single high-performance system and only prediction is distributed. This can be distributed across different edge nodes for distributed training. This can also require development of other modules for ensembling or combining the results from different trained neural networks. We will also extend this framework to make it generic so that it can be used for other deep learning applications in domains such as healthcare, agriculture and weather forecasting.

The EdgeLens software is open-source and the code sources and datasets are available at the GitHub repository https://github.com/Cloudslab/EdgeLens.

ACKNOWLEDGEMENTS

This work is supported by Melbourne-Chindia Cloud Computing Research Network and Australian Research Council. We would also like to thank Shashikant Ilager